\documentclass{iau} 
\usepackage{graphicx}

\title[JD 11.~~Gas around M33]
{The Structure of Halo Gas around M33}

\author[Olivia C. Keenan, Jonathan I. Davies, Rhys Taylor \& Robert F. Minchin]  
{Olivia C. Keenan$^1$, Jonathan I. Davies$^1$,  Rhys Taylor$^2$
 \and Robert F. Minchin$^3$}

\affiliation{$^1$School of Physics and Astronomy, Cardiff University, Queens Buildings, The Parade, Cardiff, CF24 3AA, U.K. \\ email: {\tt olivia.keenan@astro.cf.ac.uk} \\[\affilskip]
$^2$Astronomical Institute of the ASCR, Bo\v{c}n\'{i} II 1401, 14100, Prague, Czech Republic\\
$^3$Arecibo Observatory, HC03 Box 53995, Arecibo, Puerto Rico 00612}

\pubyear{2016}
\volume{321}  
\jname{Title of your IAU Symposium}
\editors{A.C. Editor, B.D. Editor \& C.E. Editor, eds.}
\begin{document}

\maketitle

\begin{abstract}
Understanding the distribution of gas in and around galaxies is vital for our interpretation of galaxy formation and evolution. As part of the Arecibo Galaxy Environment Survey (AGES) we have observed the neutral hydrogen (HI) gas in and around the nearby Local Group galaxy M33 to a greater depth than previous observations.  As part of this project we investigated the absence of optically detected dwarf galaxies in its neighbourhood, which is contrary to predictions of galaxy formation models. We observed 22 discrete clouds, 11 of which were previously undetected and none of which have optically detected counterparts.  We find one particularly interesting hydrogen cloud, which has many similar characteristics to hydrogen distributed in the disk of a galaxy. This cloud, if it is at the distance of M33, has a HI mass of around 10$^{7}$ M$_{\odot}$ and a diameter of 18 kpc, making it larger in size than M33 itself. 
\end{abstract}

\firstsection
              
                            .
\section{Introduction}

\begin{figure}[h]
\begin{center}
\includegraphics[width=3.8in]{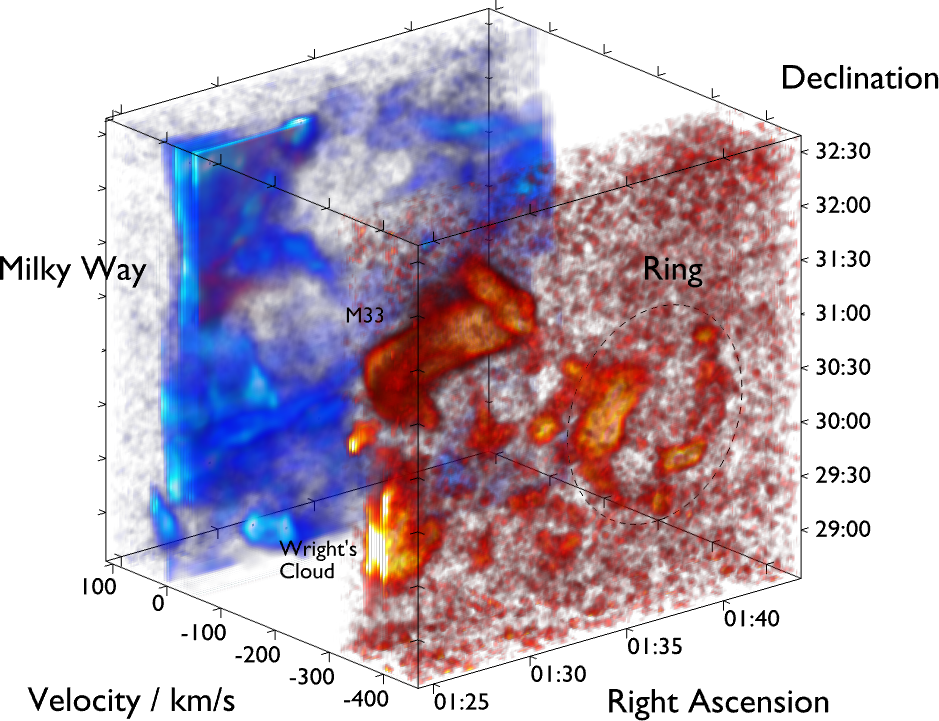} 
 \caption{Position-Velocity plot of the AGES HI data cube. The Milky Way HI is shown in blue, and everything else in red. Some of the HI structures mentioned on the text are labelled in the image. \cite[Keenan et al. 2016]{Keenan_etal16}.}
   \label{fig1}
\end{center}
\end{figure}

The dwarf galaxy problem is a discrepancy between the number of dark matter halos predicted by  $\Lambda$CDM computer simulations and the number of dwarf galaxies observed. The simulations massively over predict the amount of dwarf galaxies, when compared to observations. There are a number of suggested solutions to this problem which include: 1. That many of the missing dwarfs have very low surface brightnesses and are currently below our detection threshold. 2. That these dwarfs are extremely dark matter dominated, and some may in fact be effectively empty halos, or have gas but no stars, 3. That the simulations do not currently incorporate enough baryonic physics and, 4. $\Lambda$CDM is flawed.

The Arecibo Galaxy Environment Survey (AGES) is a HI survey covering 200 degrees of the sky (\cite[Auld et al. 2006]{Auld_etal06}). It focuses on a range of different galactic environments including clusters, isolated galaxies and local galaxies. We have used data from the survey to look at the neutral hydrogen distribution around the Local Group galaxy M33. We aim to investigate this area and search for any HI clouds which could be starless dwarf satellites of M33.

M33 currently has no confirmed dwarf satellites. It is close to the much larger M31 which has a large satellite population and dominates the gravitational potential in the area. \cite[Martin et al. 2009]{Martin_etal09} recently detected two new satellites in the M31/33 region: Andromeda XXI and Andromeda XXII. Of these Andromeda XXII's parent galaxy is ambiguous, and it has been suggested than it may be the first discovered dwarf satellite of M33 (e.g.\cite[Chapman et al. 2013]{Chapman_etal13}). There have also been previous HI observations in the region of M33. \cite[Braun \& Thilker 2004]{BraunThilker04} carried out a survey of the area using the Westerbork array and detected one cloud in the region of M33, reaching a 3$\sigma$ column density of 1.5 x 10$^{17}$ cm$^{-2}$ over 30 kms$^{-1}$ . \cite[Grossi et al. 2008]{Grossi_etal08} used data from the Arecibo Legacy Fast ALFA (ALFALFA) survey and deeper Arecibo pointings to map a 3$^\circ$ x 3$^\circ$ area around M33. They found 21 clouds in the region of M33, reaching a 1$\sigma$ column density of 5 x 10$^{17}$ cm$^{-2}$ over 10 kms$^{-1}$. Our survey extends both the column density sensitivity and spatial resolution of previous work. We reach a 1$\sigma$ column density of 1.5 x 10$^{17}$ cm$^{-2}$ over 10 kms$^{-2}$ and our resolution is 14 times better than that of \cite[Braun \& Thilker 2004]{BraunThilker04}.

\section{Results}

\begin{figure}[h]
\begin{center}
\includegraphics[width=3.4in]{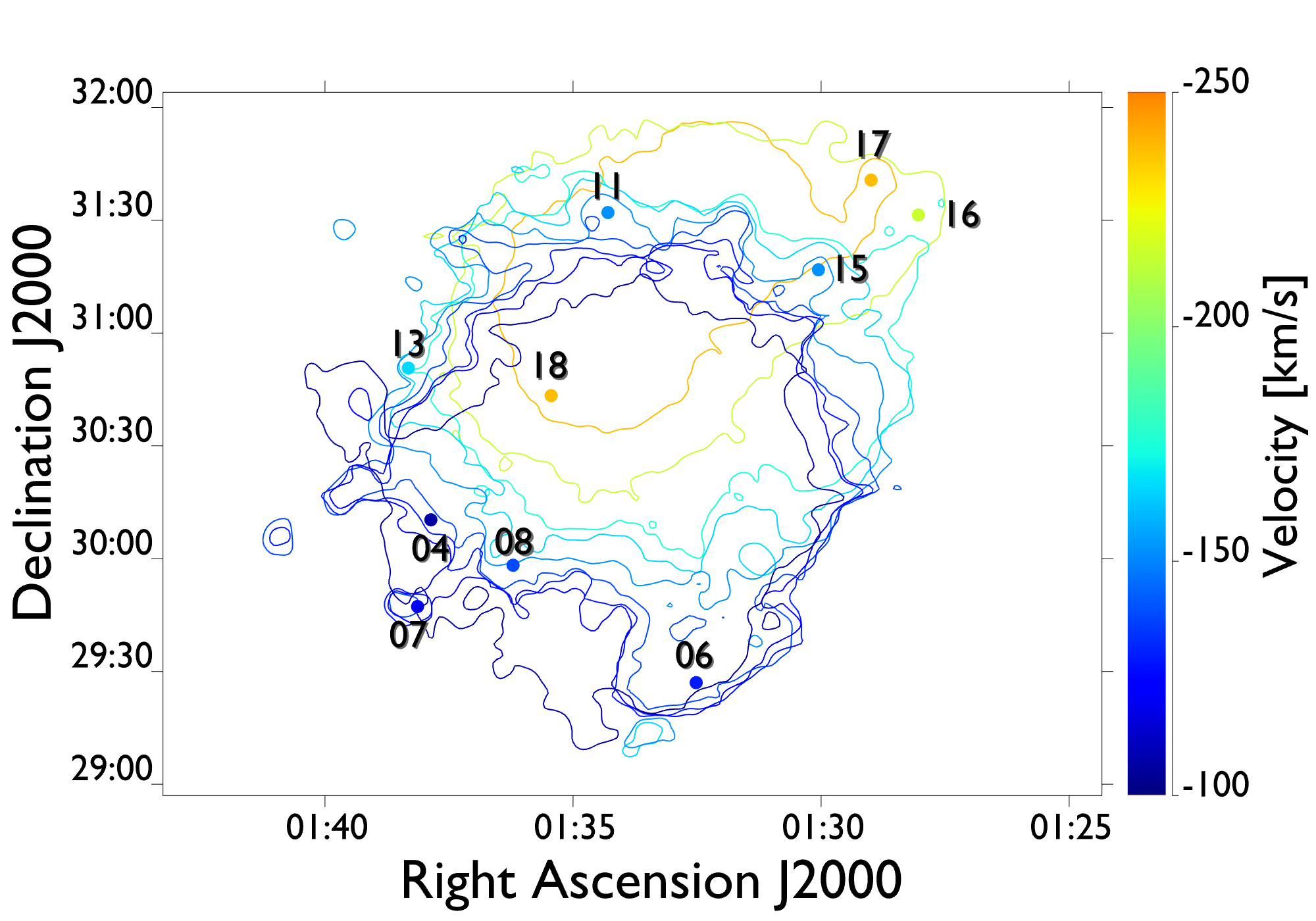} 
 \caption{A Renzogram of M33 across a velocity range of 150kms$^{-1}$. Labelled dots indicate the positions of clouds which appear to be embedded in the disk of M33. Colours are indicative of velocity. \cite[Keenan et al. 2016]{Keenan_etal16}.}
   \label{fig2}
\end{center}
\end{figure}

\begin{figure}[h]
\begin{center}
\includegraphics[width=3.4in]{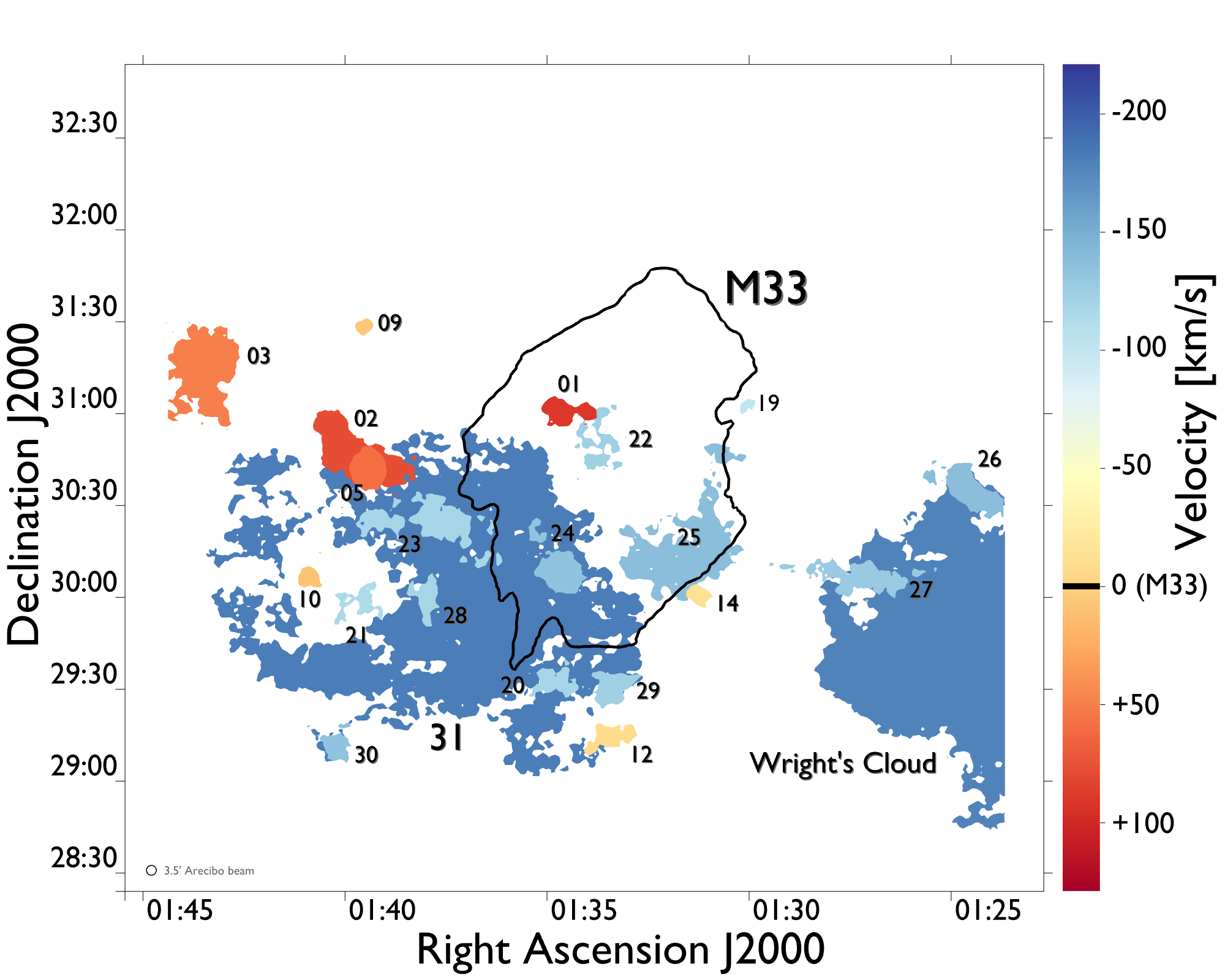} 
 \caption{The outline of M33’s HI disk is shown in black, with all of the discrete detected clouds overlaid. The figure was made by integrating the flux over the velocity range of each cloud, then filling in the lowest contour. The central velocity of each cloud was used so the colours are indicative of this velocity. Velocities shown are relative to M33. \cite[Keenan et al. 2016]{Keenan_etal16}.}
   \label{fig3}
\end{center}
\end{figure}

We detect a total of 32 clouds, 11 of which are previously undetected. Additionally, our superior column density sensitivity allows us to determine that many previously detected clouds are actually part of the low HI disk of M33. Our data cube is shown in figure \ref{fig1}, displaying the RA, Dec and velocity axes for the region containing M33. The HI emission coloured blue is from the Milky Way. When identifying clouds we had to introduce a velocity cut-off to separate Milky Way gas from structure associated with M33. Between the velocities of 0 kms$^{-1}$ and -75kms$^{-1}$ we decided that we could not distinguish Milky Way emission from that associated with M33 so we could not detect any sources in this region. Figure \ref{fig1} has M33 labelled, as well as a couple of the larger clouds within the data cube. 

Figure \ref{fig2} shows the HI disk of M33 at a range of velocities (velocity is denoted by colour). The labelled dots show clouds which fall within the HI disk at their respective velocities. These are all clouds which were initially detected by \cite[Grossi et al. 2008]{Grossi_etal08}. Figure \ref{fig3} shows all of our discrete cloud detections: those which, down to our sensitivity limit, appear separated from the disk of M33. The colour of each cloud indicates its velocity and the outline of the HI disk of M33 is shown in black. 

Our clouds have HI masses ranging from 1.0 x 10$^{5}$M$_{\odot}$ to 4.5 x 10$^{7}$M$_{\odot}$ and have velocity widths from $\sim$17 kms$^{-1}$ to $\sim$72 kms$^{-1}$. We measure the smallest of the clouds to be 1.2 kpc in size at the distance of M33. Due to the Arecibo beam size being 1kpc on the sky at this distance we wouldn't expect to measure any clouds smaller than this. All of the clouds we detect are resolved by the Arecibo beam. We have not been able to identify any stars associated with our clouds.

\section{Discussion and Conclusions}

\cite[Klyin at al. 1999]{Klypin_etal99} use CDM numerical simulations to model dark matter sub-halos around galaxies. \cite[Sternberg, McKee \& Wolfire 2002]{SternbergMckeeWolfire02} parametrise these results into an equation which allows us to predict the number of subhalos around a galaxy of a given mass. For M33 this equation predicts 25 subhalos, which is comparable to the 22 discrete clouds we observe. Our clouds also agree well with the model on the distribution of subhalo velocity widths (for full details see \cite[Keenan et al. 2016]{Keenan_etal16}).

We made moment 1 (velocity) maps of all of our clouds to look for signs of rotation. This, and the flocculent appearance of most of the clouds, leads us to conclude that they are unlikely to be bound as may have been expected if they were gas-dominated dwarf satellites.

We find one particularly interesting cloud: AGESM33-31, it is shown on the left hand side of figure \ref{fig3}. As can be seen from the figure this cloud appears to be as large as M33 and has a ring-like shape. If at the same distance as M33 it has a diameter of 18 kpc and a HI mass of 1.22 x 10$^{7}$ M$_{\odot}$. It was previously detected, but not resolved, by \cite[Thilker, Braun \& Walterbos 2002]{ThilkerBraunWalterbos02} and designated M33CHVC. It lies at the same velocity as Wright's cloud: a previously detected HI cloud part of which included in our cube (it can be seen on the right of figure \ref{fig3}). 

There are multiple proposed explanations for AGESM33-31. The first is that it is part of the Magellanic Stream. It has been suggested that Wright's cloud is part of the stream and, as AGESM33-31 has a similar velocity, it could simply be a further extension. However, the Magellanic Stream at this point of the sky is extremely sparse and made up largely of very small clouds so AGESM33-31 appears incongruent. Secondly, AGESM33-31 could be the remnants of a dark galaxy that has been disrupted through an event such as an interaction. \cite[Thilker, Braun \& Walterbos 2002]{ThilkerBraunWalterbos02} describe M33CHVC as a `dark companion' to M33. Its size, velocity width, and dispersion are comparable to that of a face on galaxy. However, these scenarios do not account for the hole in AGESM33-31. We have investigated the possibility that this hole may have been formed by a supernova, but found it to be around an order of magnitude too large for this to be a satisfactory explanation. AGESM33-31 remains an interesting and intriguing object, we would need additional observations to allow us to make further comment on its nature.

\end{document}